\documentstyle[prb,aps,epsbox,epsf]{revtex}

\begin{document} 
\draft

\title{Test of  Bell's Inequality using the Spin Filter Effect in Ferromagnetic Semiconductor Micro-structures}
\author{Shiro Kawabata\thanks {E-mail: shiro@etl.go.jp}}
\address{
Electrotechnical Laboratory, National Institute of Advanced Industrial Science and Technology, 1-1-4 Umezono, Tsukuba, 
Ibaraki 305-8568, Japan
}
%
\date{\today}
\maketitle
\begin{abstract} 

A theoretical proposal for testing Bell's inequality in mesoscopic systems is presented.
We show that the entanglement of two electron spins can be detected in the spin filter effect in the mesoscopic semiconductor / ferromagnetic semiconductor / semiconductor junction.
The current-current correlation function is calculated by use of the quantum scattering theory and we compare it with the local hidden variable theory.
We also discuss the influence of an imperfect spin filter and derive the condition to see the violation of Bell's inequality experimentally.\\

\end{abstract}
%
%
%
%
%
%
%
%
%
%

Entanglement, or quantum nonlocality between quantum systems, is  a remarkable feature of quantum mechanics which gives rise to striking phenomena such as the violation of Bell's inequality.~\cite{rf:Bell1,rf:Bell2}
Bell's inequality has already been tested experimentally with photons, i.e., massless particles.~\cite{rf:Bellexp1,rf:Bellexp2,rf:Bellexp3}
To date, however, none of experiment have been seen for massive particles such as $electrons$.
On the other hand, the semiconductor micro-fabrication techniques have allowed us to test the foundations of quantum mechanics in mesoscopic systems.
Recent experimental studies include  the fermionic two-particle interferometry (electron anti-bunching experiment)~\cite{rf:tpi} and the Hanbury Brown and Twiss experiment~\cite{rf:HBT1,rf:HBT2} in semiconductor micro-structures.
However these phenomena are not based on the nature of entangled particles.
Although many methods to generate a spin-entangled electron pair in mesoscopic devices have been proposed,~\cite{rf:entangle1,rf:entangle2,rf:entangle3,rf:entangle4,rf:entangle5} there exists no clear theoretical proposal for the test of the violation of Bell's inequality for spin-entangled electrons in these systems.

In this letter we shall show that mesoscopic semiconductor / ferromagnetic semiconductor / semiconductor (S/FS/S) systems provide a possibility to test Bell's inequality.
It was shown that the spin decoherence time for electrons in semiconductors is very long, i.e., on the order of microseconds.~\cite{rf:Kikkawa}
Therefore, the electron spin in these systems becomes a good candidate for investigating Bell's inequality in a solid state environment.
The scheme we proposed here consists of an entangler and the S/FS/S junction which act as a spin-polarized beam splitter (SPBS).
We shall show how to generate, manipulate and detect spin-entangled states experimentally.
For this purpose, we calculate the current-current correlation function by use of the quantum scattering theory and compare it with the result of a local hidden variable (LHV) theory,~\cite{rf:Bell2,rf:LHV} i.e., Bell's inequality.
We also discuss the effect  of imperfection of the SPBS in order to make a clear comparison with experiments in the future.

In the following we propose a setup which involves the entangler and two S/FS/S junctions, see Fig. 1.
The entangler is a device that produces pairs of electrons in a entangled spin singlet.~\cite{rf:entangler}.
This device can be realized in  coupled-semiconductor quantum dots, each of which contains a single electron spin.~\cite{rf:quantumdot1,rf:quantumdot2,rf:quantumdot3,rf:quantumdot4,rf:quantumdot5}
The key element of this proposal is the SPBS which ensures the spin up (down) electron leaving the entangler to be transmitted (reflected), i.e., the spin filter effect.~\cite{rf:spinfilter1,rf:spinfilter2}
It was shown that ferromagnetic semiconductors, in particular EuS~\cite{rf:EuS1,rf:EuS2} and EuSe,~\cite{rf:EuSe} can be grown in thin films which exhibit the strong spin filter effect.
In Fig. 2 we show schematic energy band diagrams for the S/FS/S junction.
Because of the exchange splitting of the electron barrier in the FS layer, spin up electron will tunnel through the barrier easily while spin down will not.
In favorable cases of EuSe, the spin polarization $
P
\equiv
(T_{\uparrow}-T_{\downarrow})/(T_{\uparrow}+T_{\downarrow})
$
($T_{\uparrow(\downarrow)}$ is the transmission probability for spin up (down) electrons)  in tunneling has exceeded $97\%$.~\cite{rf:EuSe}
In order to test Bell's inequality, we must change the relative polarization direction $\theta_2-\theta_1$ between the two FS layers arbitrary.
This can be easily achieved by magnetizing two layers along different direction before measurement.

By extending the standard quantum scattering theory,~\cite{rf:entangler,rf:scattering1,rf:scattering2} we shall calculate  the current-current correlation function for scattering with the entangled incident spin singlet state,
\begin{eqnarray}
\left|  \psi \right> 
=
\frac{1}{\sqrt{2}} 
\left[  
            a^{\dagger}_{1 \uparrow} \left( E_1 \right)  a^{\dagger}_{2 \downarrow} \left( E_2 \right)
			-
			 a^{\dagger}_{1 \downarrow} \left( E_1 \right)  a^{\dagger}_{2 \uparrow} \left( E_2 \right)
\right]
\left|  0 \right>
,
  \label{eqn:e1}
\end{eqnarray}
where $\left|  0 \right>$ denotes the filled Fermi sea, i.e., the electron ground state in the leads and $a^{\dagger}_{\alpha \sigma} \left( E \right)$ creates an incoming electron in input leads $\alpha(=1,2)$ with spin $\sigma$ (along the $z$ direction) and energy $E$.
Note that since we deal with discrete energy states here, we normalize the operator $a_{\alpha \sigma} \left( E \right)$ such that 
%
%
%
%
\begin{center}
\begin{figure}[t]
\vspace{1.0cm}
\hspace{0.0cm}
\epsfxsize=8.5cm
\epsfbox{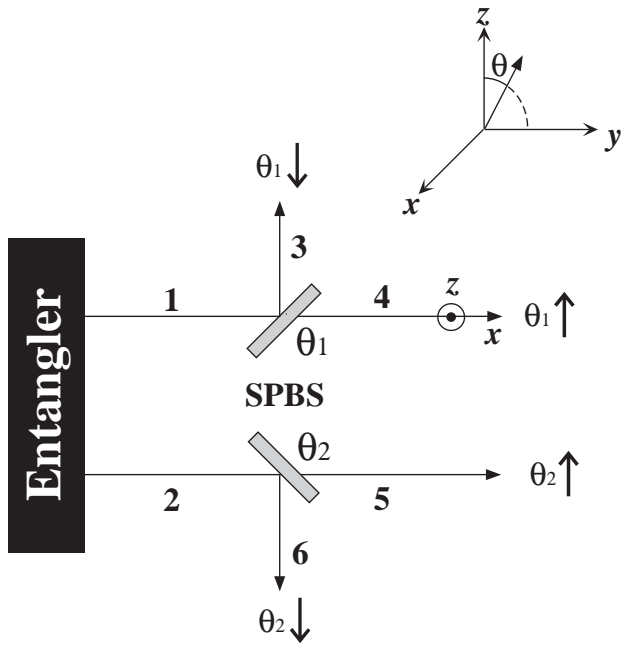}
\caption{Schematic setup for a test of the violation of Bell's inequality in mesoscopic systems.
Spin-entangled electrons generated from the entangler are fed into the spin-polarized beam splitters (SPBS) with the polarization angle $\theta_{1(2)}$ through the lead wires $\alpha=1,2$.
The entanglement of spin singlet can be detected in a current correlation measurement at the output leads ($\alpha=3,4,5,6)$.
}
\end{figure}
\end{center}
%
%
%
%
%
%
%
%
%
\begin{center}
\begin{figure}[t]
\vspace{1.0cm}
\hspace{-0.5cm}
\epsfxsize=7.0cm
\epsfbox{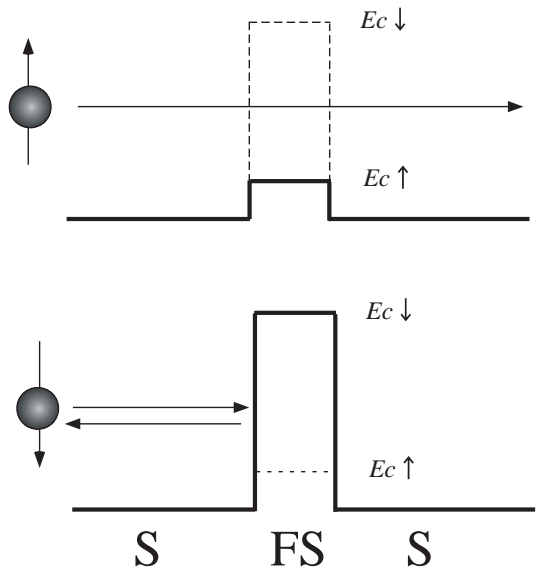}
\caption{Conduction band profile of the spin filter semiconductor / ferromagnetic semiconductor / semiconductor (S/FS/S) junction.
$E_{c\uparrow(\downarrow)}$ is the conduction band energy for spin up (down) electrons in the FS layer.
Exchange interaction gives rise to a spin-dependent potential; spin down electrons see a large barrier while spin up electron a small one.
}
\end{figure}
\end{center}
%
%
%
%
%
%
%
%
\begin{eqnarray}
\left[ 
           a_{\alpha \sigma} \left( E \right)
		   ,
		   a^{\dagger}_{\beta \sigma'} \left( E' \right)
\right]
=
\delta_{\sigma,\sigma'} \delta_{\alpha,\beta} \delta_{E,E'}/\nu
,
  \label{eqn:e2}
\end{eqnarray}
where $\nu$ is the density of states in the lead wires.
We assume that each lead wire consists of a single quantum channel.
Then the current operator in the output leads $\alpha(=3,4,5,6)$ is given by
\begin{eqnarray}
I_{\alpha \sigma}  \left( \theta, t \right)
=
\frac{e}{h\nu} \sum_{E,E'}
\left[
          a^{\dagger} _{\alpha \sigma} \left(\theta,E \right)
          a _{\alpha \sigma} \left(\theta,E' \right)
		  -
          b^{\dagger} _{\alpha \sigma} \left(\theta,E \right)
          b _{\alpha \sigma} \left( \theta,E' \right)
\right]
\exp \left[  i \frac{E-E'}{\hbar} t\right]
.
  \label{eqn:e3}
\end{eqnarray}
The operator $a^{\dagger} _{\alpha \sigma} \left(\theta,E \right)$ creates the electron with spin $\sigma$ along a specified direction with the polar angle $\theta$ (see Fig. 1). 
Then the relation between operator $a _{\alpha \sigma} \left(\theta,E \right)$ and $a _{\alpha \sigma} \left(E \right)$ is given by the spinor transformation
\begin{eqnarray}
\left(
\matrix{
     a _{\alpha \uparrow} \left(\theta ,E\right) \cr
     a _{\alpha \downarrow} \left(\theta  ,E\right)  
}
\right)
=
\left(
\matrix{
 \cos   \frac{\theta}{2}  & \sin  \frac{\theta}{2}  \cr
-\sin   \frac{\theta}{2}  & \cos   \frac{\theta}{2} 
}
\right)
\left(
\matrix{
      a _{\alpha \uparrow}  \left(E\right)\cr
     a _{\alpha \downarrow}\left(E\right)
}
\right)
.
  \label{eqn:e4}
\end{eqnarray}
In eq. (\ref{eqn:e3}), the operator $b_{\alpha \sigma} \left(\theta,E \right)$ is related to the operator $a_{\alpha \sigma} \left(\theta,E \right)$ via the scattering matrix $s_{\alpha\sigma,\beta\sigma}$,
\begin{eqnarray}
b _{\alpha \sigma} \left(\theta,E \right)
 =\sum_{\beta=1}^6 s_{\alpha\sigma,\beta\sigma}
  a _{\beta \sigma} \left(\theta,E \right)
  .
  \label{eqn:e5}
\end{eqnarray}
%
%
%
Using eq.~(\ref{eqn:e3}) and ~(\ref{eqn:e5}), we arrive at the following expression for the current operator
\begin{eqnarray}
I_{\alpha \sigma} \left( \theta, t \right)
&=&
\frac{e}{h\nu} 
\sum_{E,E'}
\sum_{\beta,\gamma=1}^6
a^{\dagger} _{\beta \sigma} \left(\theta,E \right)
A^{\alpha}_{\beta\gamma} \left( \sigma \right)
a_{\gamma \sigma} \left(\theta,E' \right)
\exp \left[  i \frac{E-E'}{\hbar} t\right]
,
  \label{eqn:e6}
\\
A^{\alpha}_{\beta\gamma} \left( \sigma \right)
&\equiv&
\delta_{\alpha,\beta}\delta_{\alpha,\gamma}
-
s_{\alpha\sigma,\beta\sigma}^{*}s_{\alpha\sigma,\gamma\sigma}
.
  \label{eqn:e7}
\end{eqnarray}
The current-current correlation function between the leads $\alpha(=3,4)$ and $\beta(=5,6)$ is given by
\begin{eqnarray}
C_{\alpha\sigma,\beta\sigma'} (\theta_1,\theta_2)
&\equiv&
\lim_{t' \rightarrow \infty}
\frac{1}{t'}
\int_{0}^{t'}
d t
\left< \psi \right|   
                I_{\alpha\sigma} \left( \theta_1,t\right)  
				I_{\beta\sigma'}  \left( \theta_2,t\right) 
\left|   \psi  \right>
\nonumber\\
&=&
\frac{e^2}{h^2\nu^2}
\sum_{E,E'}
\sum_{\gamma,\delta,\varepsilon,\xi=1}^6
A^{\alpha}_{\gamma\delta} \left( \sigma \right)
A^{\beta}_{\varepsilon\xi} \left( \sigma' \right)
\left< \psi \right|   
                a^{\dagger} _{\gamma \sigma} \left(\theta_1,E \right)
				a _{\delta \sigma} \left(\theta_1,E \right)
                a^{\dagger} _{\varepsilon \sigma'} \left(\theta_2,E' \right)
				a _{\xi \sigma'} \left(\theta_2,E' \right)
\left|   \psi  \right>
.
  \label{eqn:e8}
\end{eqnarray}
Substituting $\left| \psi \right>$ defined in eq.~(\ref{eqn:e1}) into eq.~(\ref{eqn:e8}) and using the commutation relation eq.~(\ref{eqn:e2}), we obtain
\begin{eqnarray}
C_{4\uparrow,5\uparrow} (\theta_1,\theta_2)
=
C_{3\downarrow,6\downarrow} (\theta_1,\theta_2)
=
\frac{e^2}{2h^2\nu^2} 
\sin^2 \left( \frac{\theta_1-\theta_2}{2} \right)
,
  \label{eqn:e9}
\\
C_{4\uparrow,6\downarrow} (\theta_1,\theta_2)
=
C_{3\downarrow,5\uparrow} (\theta_1,\theta_2)
=
\frac{e^2}{2h^2\nu^2} 
\cos^2 \left( \frac{\theta_1-\theta_2}{2} \right)
.
  \label{eqn:e10}
\end{eqnarray}
In  order to compare with the LHV theory,~\cite{rf:LHV} we consider the following function,
\begin{eqnarray}
F \left( \theta_1,\theta_2\right)
\equiv
\frac{
           \displaystyle{
           C_{4\uparrow,5\uparrow} (\theta_1,\theta_2)
		   +
		   C_{3\downarrow,6\downarrow} (\theta_1,\theta_2)
		   -
		   C_{4\uparrow,6\downarrow} (\theta_1,\theta_2)
		   -
		   C_{3\downarrow,5\uparrow} (\theta_1,\theta_2)
          }
		  }
		  {
          \displaystyle{
           C_{4\uparrow,5\uparrow} (\theta_1,\theta_2)
		   +
		   C_{3\downarrow,6\downarrow} (\theta_1,\theta_2)
		   +
		   C_{4\uparrow,6\downarrow} (\theta_1,\theta_2)
		   +
		   C_{3\downarrow,5\uparrow} (\theta_1,\theta_2)
		  }
		  }
		  .
  \label{eqn:e11}
\end{eqnarray}
Assuming that the two SPBS are ideal (
$
T_\uparrow
\equiv
\left| s_{4\uparrow,1\uparrow}\right|^2
=
\left| s_{5\uparrow,2\uparrow}\right|^2=1
$
, 
$
R_\downarrow
\equiv
\left| s_{3\downarrow,1\downarrow}\right|^2
=
\left| s_{6\downarrow,2\downarrow}\right|^2=1
$
),  we  obtain 
\begin{eqnarray}
F \left( \theta_1,\theta_2\right)
		  =
		  -\cos \left( \frac{\theta_1-\theta_2}{2} \right)
		  .
  \label{eqn:e12}
\end{eqnarray}
Bell showed that any LHV theory~\cite{rf:LHV} must obey the inequality,
\begin{eqnarray}
\left|     B \left( \theta_1,\theta_2,\theta'_1,\theta'_2\right)   \right|
\equiv
\left|  
           F \left( \theta_1,\theta_2\right) 
		- F \left( \theta_1,\theta'_2\right)  
		+ F \left( \theta'_1,\theta_2\right)
		+ F \left( \theta'_1,\theta'_2\right)    
\right|
\le 2
.
   \label{eqn:e13}
\end{eqnarray}
This inequality is called Bell's inequality.
On the other hand, our quantum mechanical calculation gives the result
\begin{eqnarray}
B \left( \theta_1,\theta_2,\theta'_1,\theta'_2\right)
=
-\cos \left( \frac{\theta_1-\theta_2}{2} \right)
+\cos \left( \frac{\theta_1-\theta'_2}{2} \right)
-\cos \left( \frac{\theta'_1-\theta'_2}{2} \right)
-\cos \left( \frac{\theta'_1-\theta_2}{2} \right)
.
  \label{eqn:e14}
\end{eqnarray}
By choosing, $\theta_1-\theta_2=\theta'_1-\theta'_2=\theta'_1-\theta_2=(\theta_1-\theta'_2)/3\equiv\Theta$, one gets 
\begin{eqnarray}
B \left( \Theta \right)
=
\cos 3 \Theta -3 \cos \Theta
.
  \label{eqn:e15}
\end{eqnarray}
When $\Theta=3\pi/8$, $B=2\sqrt{2}$ showing clear violation of Bell's inequality $\left| B \right| \le2$ (see Fig. 2).
%
%
%
%
\begin{center}
\begin{figure}[t]
\vspace{1.0cm}
\hspace{-0.5cm}
\epsfxsize=8.0cm
\epsfbox{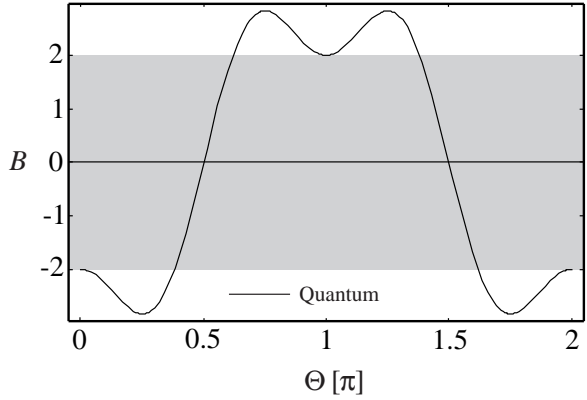}
\caption{The quantum correlation (solid line) as a function of the relative angle $\Theta$ of the spin-polarized beam splitters and the region which satisfy the Bell's inequality (shaded area).}
\end{figure}
\end{center}
%
%
%
%
%
%

In the above calculation, we have assumed that the SPBS is ideal, i.e., the transmission probability $T_{\uparrow}$ for spin up electrons equals to one, and the reflection probability $R_{\downarrow}$ for spin down electrons also equals to one.
However, in practice it is difficult to fabricate a perfect SPBS, because the electron affinity of the S layer is generally different from that of the FS layer.~\cite{rf:EuS1}
This gives rise to imperfect transmission for spin up electrons, i.e., $T_{\uparrow}<1$.
Moreover, as a result of a finite barrier height for spin down electrons in real junctions,  $R_{\downarrow}$ becomes less than unity.
Therefor, it is important to see how small the values $T_{\uparrow}$ and $R_{\downarrow}$ can be to have still the violation of  Bell's inequality, i.e., $\left| B \right|>2$.
Below we shall derive the condition for $T_{\uparrow}$ and $R_{\downarrow}$ by extending the above calculation.

In the case of a imperfect SPBS, the current operator in each output leads is expressed as a sum of a contribution from spin up and spin down electrons, i.e,  $I_{\alpha}(\theta,t) = \sum_{\sigma=\uparrow,\downarrow}I_{\alpha,\sigma}(\theta,t)$.
By calculating the current-current correlation functions,
$
C_{\alpha,\beta} (\theta_1,\theta_2)
\equiv
$
$
\left< \psi \right|   
                I_{\alpha} \left( \theta_1,t\right)  
				I_{\beta}  \left( \theta_2,t\right) 
\left|   \psi  \right>,
$
and using eq.~(\ref{eqn:e11}), we obtain 
\begin{eqnarray}
B (\Theta)
&=&
\frac{1}{2}
\left[
             \left\{
			                 \left( T_{\uparrow} - R_{\uparrow}  \right)^2 
							 +
			                 \left( T_{\downarrow} - R_{\downarrow}\right)^2 
			 \right\}
\left(
          3 \sin^2  \frac{\Theta}{2} - \sin^2 \frac{3\Theta}{2}
\right)
+
			 2
			                 \left( T_{\uparrow} - R_{\uparrow}  \right)
			                 \left( T_{\downarrow} - R_{\downarrow}\right)
\left(
          3 \cos^2  \frac{\Theta}{2} - \cos^2 \frac{3\Theta}{2}
\right)
\right]
.
  \label{eqn:e20}
\end{eqnarray}
Therefore, to have the violation of Bell's inequality, i.e., $B>2$ at $\Theta=3\pi/8$, $T_{\uparrow}$ and $R_{\downarrow}$ must satisfy following condition:
\begin{eqnarray}
\frac{c}{2}
\left[
(2 T_{\uparrow} -1)^2
+
(2 R_{\downarrow} -1)^2
\right]
+
\frac{c-2}{2}
(2 T_{\uparrow} -1)
(2 R_{\downarrow} -1)
>
1
,
  \label{eqn:e21}
\end{eqnarray}
where $c \equiv 3 \sin^2  (3\pi/8) - \sin^2 (9\pi/3)$ and we have used $T_{\downarrow}=1-R_{\downarrow}$ and $R_{\uparrow}=1-T_{\uparrow}$.
Fig. 4 shows the region which is forbidden by the Bell's inequality, i.e., eq.~(\ref{eqn:e13}).
Therefore we must use the S/FS/S junctions which satisfy the condition eq.~(\ref{eqn:e21}) to see experimentally the violation of Bell's inequality.
Especially in the case of $T_\uparrow=R_\downarrow$, the condition for $B>2$ is given by
\begin{eqnarray}
T_\uparrow=R_\downarrow
>
\frac{1}{2} + \frac{1}{2\sqrt{c-1}}
\approx
0.92
.
  \label{eqn:e22}
\end{eqnarray}
This corresponds to the  spin polarization $P >84\%$.
Such spin filter effect can be realized with nowadays spin electronics technology~\cite{rf:EuSe}.

In conclusion, by measuring the current-current correlation in the mesoscopic system described here, one can probe the entanglement of electron spins.
This can be used as an experimental test of the violation of Bell's inequality.
Spin-entangled electrons produced by the entangler, i.e., coupled quantum dots, enter the SPBS, i.e., the spin filter S/FS/S junctions.
We have calculated the function $B$ for this setup by using the quantum mechanical scattering theory and compare it with the LHV theory.
Moreover, to compare with experiments in the future we have also derived the condition for testing the violations of Bell's inequality in the case of the imperfect SPBS.

The device setup described here will be of immediate use in further experimental tests on the foundations of quantum mechanics (e.g., Greenberger-Horne-Zeilinger correlation~\cite{rf:GHZ1,rf:GHZ2}, quantum teleportation~\cite{rf:teleportation} and quantum eraser experiment~\cite{rf:eraser}) and on the quantum information technology (e.g., quantum computation~\cite{rf:q-comp} and quantum cryptography~\cite{rf:crypto}) in $mesoscopic$ condensed matter systems. 
%
%
%
%
\begin{center}
\begin{figure}[t]
\vspace{1.0cm}
\hspace{-0.5cm}
\epsfxsize=9.0cm
\epsfbox{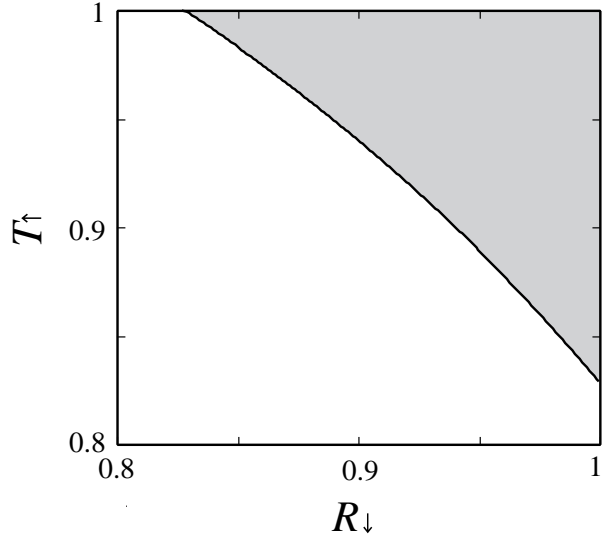}
\caption{Diagram of the condition for $T_{\uparrow}$ and $R_{\downarrow}$ (eq.~(\ref{eqn:e21})) in the case of $\Theta=3\pi/8$.
The shaded region is forbidden by the Bell's inequality.}
\end{figure}
\end{center}
%
%
%
%
%
%
%
%
%
%
%
%
%
%
%
%

The author is grateful to
S. Abe, K. Ando, J. Bird, Y. Suzuki, E. Tamura, and S. Yuasa for useful discussions.

\end{document}